\documentstyle[aps,prl,epsfig,multicol,times]{revtex}
\makeatletter
\newlength\abovecaptionskip \newlength\belowcaptionskip
\def\@makecaption#1#2{%
 \vskip\abovecaptionskip \sbox\@tempboxa{#1: #2}%
 \ifdim \wd\@tempboxa >\hsize #1: #2\par \else \global \@minipagefalse
 \hb@xt@\hsize{\hfil\box\@tempboxa\hfil}%
 \fi \vskip\belowcaptionskip} \makeatother
\begin{document}
\title{No indications of metal-insulator transition 
for systems of interacting electrons \\ in two dimensions} 
\author{Richard Berkovits$^{1,2,3}$, Jan W. Kantelhardt$^{1,4}$, 
Yshai Avishai$^{1}$, Shlomo Havlin$^{1}$, 
Armin Bunde$^{4,1}$}
\address{$^{1}$ Minerva Center and Department of Physics,
Bar-Ilan University, Ramat-Gan 52900, Israel}
\address{$^{2}$ Physics Department,
Princeton University, Princeton, NJ 08544}
\address{$^{3}$ NEC Research Institute, 4 Independence Way,
Princeton, NJ 08540}
\address{$^{4}$ Institut f\"ur Theoretische Physik III, 
Universit\"at Giessen, D-35392 Giessen, Germany}
\date{June 9, 2000, version 6.2}
\draft
\maketitle
\begin{multicols}{2}[%
\begin{abstract} 
The influence of Coulomb interaction on transport properties of 
spinless electrons in small disordered two dimensional systems 
is studied within a tight binding model. Spatial correlations, 
inverse participation ratio, and multifractal spectrum of the zero 
temperature local tunneling amplitude as well as the DC Kubo
conductance are traced as function of the interaction strength 
$U$. When $U$ is increased, all of the above quantities
are shifted rather smoothly
towards localized behavior, indicating the 
absence of an interaction driven insulator-metal transition.
\end{abstract}
]

Recently, much interest is focused on the influence of electron-electron 
interaction ({\em eei})
on the localization properties of two dimensional electrons
 in disordered systems. 
It is motivated by the experimental observations of a crossover
in the behavior of the conductance of low
density two dimensional electrons from an insulating like temperature 
dependence at low densities to a metallic one at higher
densities \cite{kravchenko94}. In some cases a transition back to an 
insulating dependence at even higher densities was 
observed \cite{hamilton99}.  This so dubbed 
2DMIT (two dimensional
``metal-insulator'' transition) is at odd with the
scaling theory of localization \cite{abrahams79} which, 
for non-interacting electrons, asserts that
all states in 2D are localized by any amount of disorder.
It prompted an intensive theoretical effort including
analytical \cite{dobrosavljevic97} and numerical 
\cite{talamantes96,cuvas99,benenti99,waintal99,shepelyansky99,denteneer99} 
work which tried to explain the 2DMIT as a result of {\em eei},
not taken into account by the scaling theory. Another viewpoint,
supported by some recent experiments \cite{papadakis99}, 
maintains that there is no 
2DMIT (that is, the systems are insulating at zero 
temperature) and that the origin of the
observed temperature dependence 
is not a result of a metallic zero temperature phase
\cite{altshuler99}.

In the absence of {\em eei},
there are well established relations
between the conductance and the properties of individual
single electron states. Such relations do not 
exist for the many electron states. For example,
the sensitivity of a single electron state at the Fermi energy to 
a change in the boundary condition is a measure of the
conductance of the system (the Thouless criteria \cite{edwards72}).
Formulation of a similar conjecture for many-electron states
is much more intricate \cite{berkovits96,akkermans97}.
Another example is the behavior of the level statistics at
the Anderson localization transition. This transition
is accompanied by a transition in
the single electron energy level statistics from Wigner to Poisson
\cite{shklovskii93}, while the many particle energy level statistics
for non-interacting electrons
will remain Poissonian both in the metallic and in the 
insulator regime \cite{berkovits94}. 

In previous studies of 2D spinless electrons with {\em eei}
\cite{talamantes96,cuvas99,benenti99,waintal99,shepelyansky99} 
numerous equilibrium properties such as
many particle energy level statistics,
two-electron localization length high above the Fermi energy, 
persistent current flow patterns,
and charge density response to an external perturbation were studied.
As function of the interaction strength $U$, the footprints
of some of these 
quantities  undergo qualitative changes. It might then be tempting 
to interpret it as an
evidence for a 2DMIT. 
However, the 
properties listed above are not directly related to
zero temperature transport properties of the electron system, which
are naturally measured in experiments.
It is therefore mandatory to
carefully examine some quantities which are 
directly related to transport properties
of the system. Since the problem under study involves disorder and 
{\em eei} the region of interest in parameter space is that for which
both of them are significant. This excludes the 
possibility of using perturbation theory, and 
leaves us with the necessity of employing exact numerical diagonalization. 

This task, which is feasible for relatively small systems,
is carried out in the present Letter. 
The statistical properties of
tunneling amplitudes and the conductance distribution
of the system are examined. 
It is shown that {\em eei} systematically
{\it attenuates} the transport through the system and 
enhances its
insulating features. Thus, although {\em eei} may significantly
alter some of the properties of 
a 2D electron system as demonstrated
in previous studies \cite{cuvas99,benenti99,waintal99,shepelyansky99}, 
it is conjectured here that
there is no numerical evidence that it can drive a 
transition in the transport properties of spinless electrons.

We consider systems composed of $4$, $6$, and $8$ interacting electrons
residing on $6 \times 6$, $5 \times 5$, and $5 \times 4$ lattices 
(correspondingly) having a torus geometry. 
It is of course not possible here 
to directly mimic the experimental 
procedure in which the conjectured 2DMIT is 
driven through variation of electron density. 
Instead, the physical content of this
density variation can be captured by 
controlling the ratio of the
Fermi energy to the interaction energy. 
In the present model, it is achieved simply
by changing the interaction strength $U$ while keeping other 
parameters intact \cite{benenti99}.

In the Tight-Binding approximation, the
Hamiltonian of the system is given by:
\begin{eqnarray}
H= \sum_{k,j} \epsilon_{k,j} a_{k,j}^{\dag} a_{k,j} - V \sum_{k,j}
(a_{k,j+1}^{\dag} a_{k,j} + h.c)\nonumber \\
- V \sum_{k,j}
(a_{k+1,j}^{\dag} a_{k,j} + h.c)+
U \sum_{k,j>l,p} {{a_{k,j}^{\dag} a_{k,j}
a_{l,p}^{\dag} a_{l,p}} \over 
{|\vec r_{k,j} - \vec r_{l,p}|/s}}
\label{hamil}
\end{eqnarray}
where $\vec r=(k,j)$ denotes a lattice site,
$a_{k,j}^{\dag}$ is an electron creation operator,
$\epsilon_{k,j}$ is the site energy, chosen 
randomly between $-W/2$ and $W/2$ with uniform probability, $V$
is a constant hopping matrix element and $s$ is the lattice constant. 
Using the Lanczos method we obtain the 
many-particle eigenvalues $\varepsilon_\alpha^N$ 
and eigenfunctions $|\alpha^N\rangle$, 
where $N$ is the number of electrons.
The zero temperature local tunneling amplitude 
$\langle 0^{N} | a_{\vec r}^{\dag} |0^{N-1}\rangle$
between the ground state of $N$ and $N-1$ electrons 
can be employed here
in order to characterize the transport
properties of the many-particle interacting system.
It has the advantage that only the ground state
energy and eigenvector for $N$ and $N-1$ electrons need to be
calculated.  The use of the tunneling 
amplitude in this context can be motivated and substantiated
by the following considerations: 
In the independent particle approximation,
the tunneling density of state (TDOS) is
given by 
\begin{eqnarray}
\nu(\varepsilon)=\sum_n \int |\psi_n(\vec r)|^2 \, dr
\; \delta(\varepsilon - \varepsilon_n) = \sum_n \delta(\varepsilon - 
\varepsilon_n),
\label{tdossingle}
\end{eqnarray}
where $\psi_n$ is the $n$-th single particle eigenvector and 
$\epsilon_n$ is the $n$-th single electron eigenvalue. 
For the many-body interacting system,
the TDOS is defined as \cite{efros95}:
\begin{eqnarray}
\nu(\varepsilon)=\sum_\alpha| 
\langle\alpha^{N} | \sum_{\vec r} a_{\vec r}^{\dag} |0^{N-1}\rangle |^2 
\; \delta(\varepsilon - (\varepsilon_{\alpha}^{N}-\varepsilon_0^{N-1})).
\label{tdos}
\end{eqnarray}
The conductance $\sigma(\varepsilon)$ is related
to the transmission $t(\vec r,\vec r ~',\varepsilon)$ of an
electron with energy $\varepsilon$ between two points $\vec r$,
$\vec r~'$ on the interface of the system with external leads
through the Landauer formula \cite{landauer57}
$\sigma(\varepsilon)= (e^2/h) \sum_{\vec r, \vec r~'} 
| t(\vec r,\vec r ~',\varepsilon) |^2 $, where the sum is over all points
on the interface. For a non-interacting system the transmission is 
expressible in terms of single particle wave functions,
\begin{eqnarray}
t(\vec r,\vec r ~',\varepsilon) = \sum_n \psi_n^*(\vec r) 
\psi_n(\vec r~') \; \delta(\varepsilon - \varepsilon_n).
\label{tsingle}
\end{eqnarray}
As discussed in Ref. \cite{meir92}, an appropriate
Landauer formula connecting transmission and conductance for an 
interacting system coupled to external leads 
has a similar structure, in which
the transmission
in the interacting region is given by
\begin{eqnarray}
t(\vec r,\vec r ~',\varepsilon) = \sum_{\alpha}
\langle \alpha^{N} | a_{\vec r}^{\dag} |0^{N-1}\rangle
\langle 0^{N-1} | a_{\vec r ~'} |\alpha^{N}\rangle \nonumber \\
\delta (\varepsilon - (\varepsilon_{\alpha}^{N}-
\varepsilon_0^{N-1})).
\label{t}
\end{eqnarray}

In the absence of interaction, Eq.~(\ref{tdos}) is reduced to
Eq.~(\ref{tdossingle}), and Eq.~(\ref{t}) becomes identical to
Eq.~(\ref{tsingle}). Indeed, the tunneling amplitude
$\langle \alpha^{N} | a_{\vec r}^{\dag} |0^{N-1}\rangle $
couples only many-particle states different from each other by
the addition of a single particle to an unoccupied single particle
state. Thus, for a many-particle state $|\alpha^{N}\rangle$ which
corresponds to a Slater determinants in which
all $N-1$ single particle state are occupied as well as
the $n$-th single particle state above the Fermi energy,
one obtains $\psi_{n}(\vec r) =
\langle \alpha^{N} | a_{\vec r}^{\dag} |0^{N-1}\rangle $ and
$\varepsilon_{\alpha}^{N}-\varepsilon_0^{N-1}
= \epsilon_{n}$, while for other many particle states the matrix element
vanishes. This behavior suggests that the tunneling amplitude is
the appropriate quantity to replace the single electron wave function
in studying transport properties of interacting systems. 
A similar procedure is employed in Ref. \cite{jeon99} in order
to generalize the concept of
inverse participation ratio for interacting systems.

Note, however, that once interactions are present
a many particle state is a superposition of many 
different Slater determinants, and, generically,
$\sum_{\vec r} |\langle 0^{N} | a_{\vec r}^{\dag} 
|0^{N-1}\rangle|^2 \ne 1$, where for states corresponding
to quasi-particles the matrix element are dominant
\cite{mahan90}. Thus, for interacting systems the tunneling amplitude
$\langle 0^{N} | a_{\vec r}^{\dag} |0^{N-1}\rangle$ is not normalized.
The reason is, that, due to interaction, the 
basic objects are
quasi-particles (rather than particles) with
a finite life time. In this context,
interaction affects transport properties of a given system
in two different ways.  The first (which is not related at all 
to quantum localization) is implied by
variation of the density of states at a 
given energy, while the second one (which is the essence of quantum 
localization) is manifested through a change
in the correlation properties of the tunneling amplitude between different
points. 
Therefore, in order to study the influence of {\em eei} on quantum
localization it is useful to define an effective tunneling amplitude 
$\phi(\vec r) = \langle 0^{N} | a_{\vec r}^{\dag} |0^{N-1}\rangle /
(\sum_{\vec r} \langle 0^{N} | a_{\vec r}^{\dag} |0^{N-1}\rangle^2)^{1/2}$, 
which will be used to study transport properties 
of an interacting system.

\begin{figure}\centering 
\epsfxsize7.5cm\epsfbox{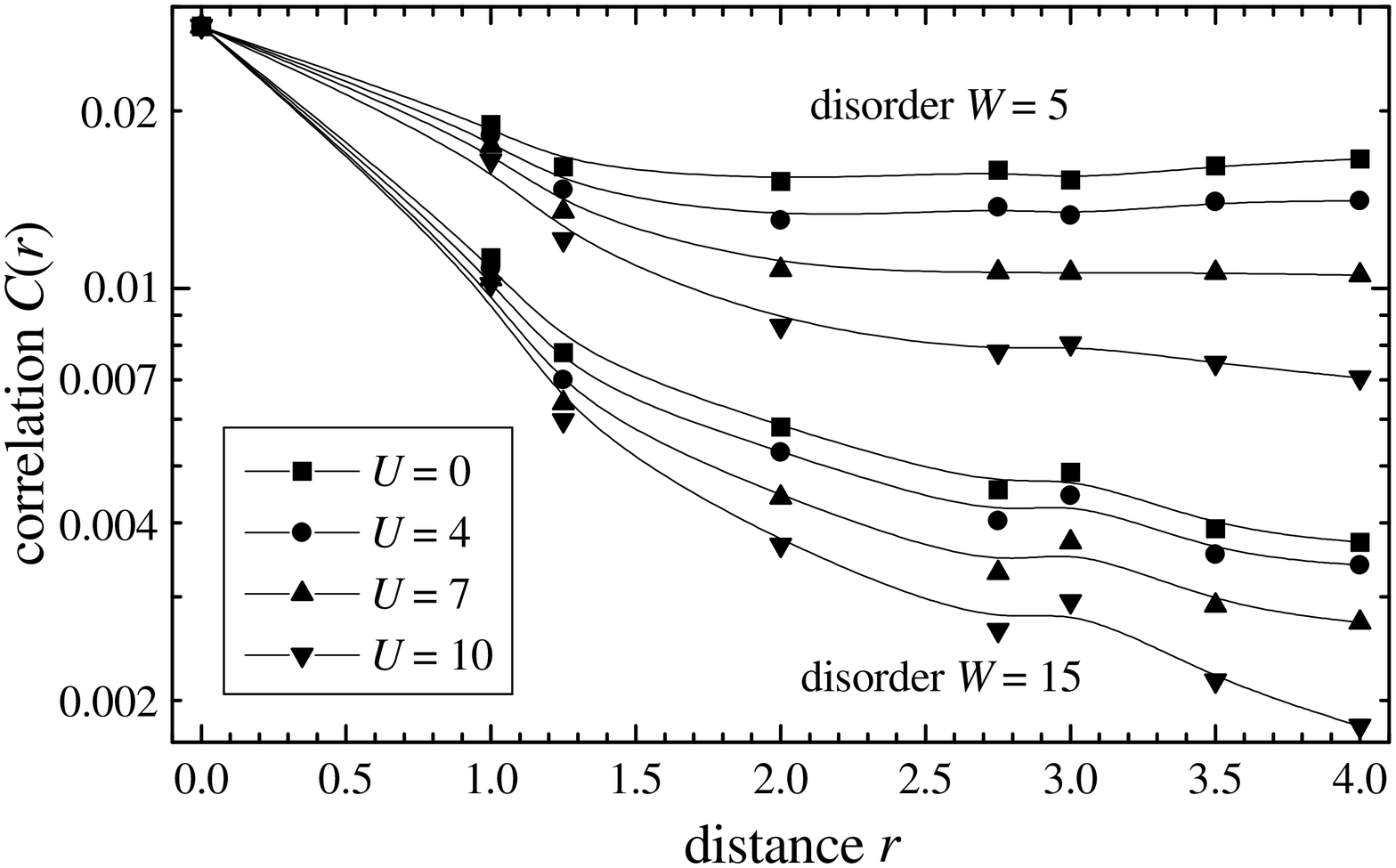}
\parbox{8.5cm}{\caption[]{\small 
The decay of the correlation function 
$C(r)$ for $\phi(\vec r)$ for two different
disorder strengths $W=5$ (in units of $V$) in the upper part and $W=15$ in the 
lower part and 4 interaction strengths (see legend).  The correlation function 
decays faster for stronger disorder and for stronger interaction.}
\label{fig:results1}}\end{figure}

The effective tunneling amplitude
$\phi(\vec r)$ of the system introduced above is calculated 
for several
values of disorder, and then analyzed by employing well
known procedures for elucidating properties of
single particle wave functions such as
spatial correlation function, inverse participation ratio and a 
multifractal analysis.
Results for the correlation function 
\begin{equation}
  C(r) = \langle \vert \phi(\vec r) \phi(\vec r~') \vert 
  \rangle_{\vert \vec r - \vec r~' \vert = r},
\end{equation}
for the $6 \times 6$, $4$ electrons system
averaged over $100$ realizations
are shown in Fig.~\ref{fig:results1}
(similar results are obtained for the other system sizes).
Evidently, $C(r)$ decays as function of $r$ indicating the loss of 
amplitude correlations.  The decay is faster for stronger disorder
as is the case for a system of 
non-interacting electrons.  Moreover,
the decay of $C(r)$ evolves smoothly with increasing 
interaction strength $U$, indicating stronger localization at 
higher values of $U$,
showing no sign of a 2DMIT or an intermediate 
metallic phase. These results are significant, since
an identical system as considered here 
undergoes a transition in the 
character of its persistent current at intermediate
values of $U$ \cite{berkovits98,benenti99}. In particular, the 
flow pattern of the local persistent current is 
shown to be ordered and essentially
one dimensional. 
Our new result indicate that this kind of behavior 
should not be interpreted as a
 signature of a delocalized (Coulomb) phase.

\begin{figure}\centering 
\epsfxsize8.5cm\epsfbox{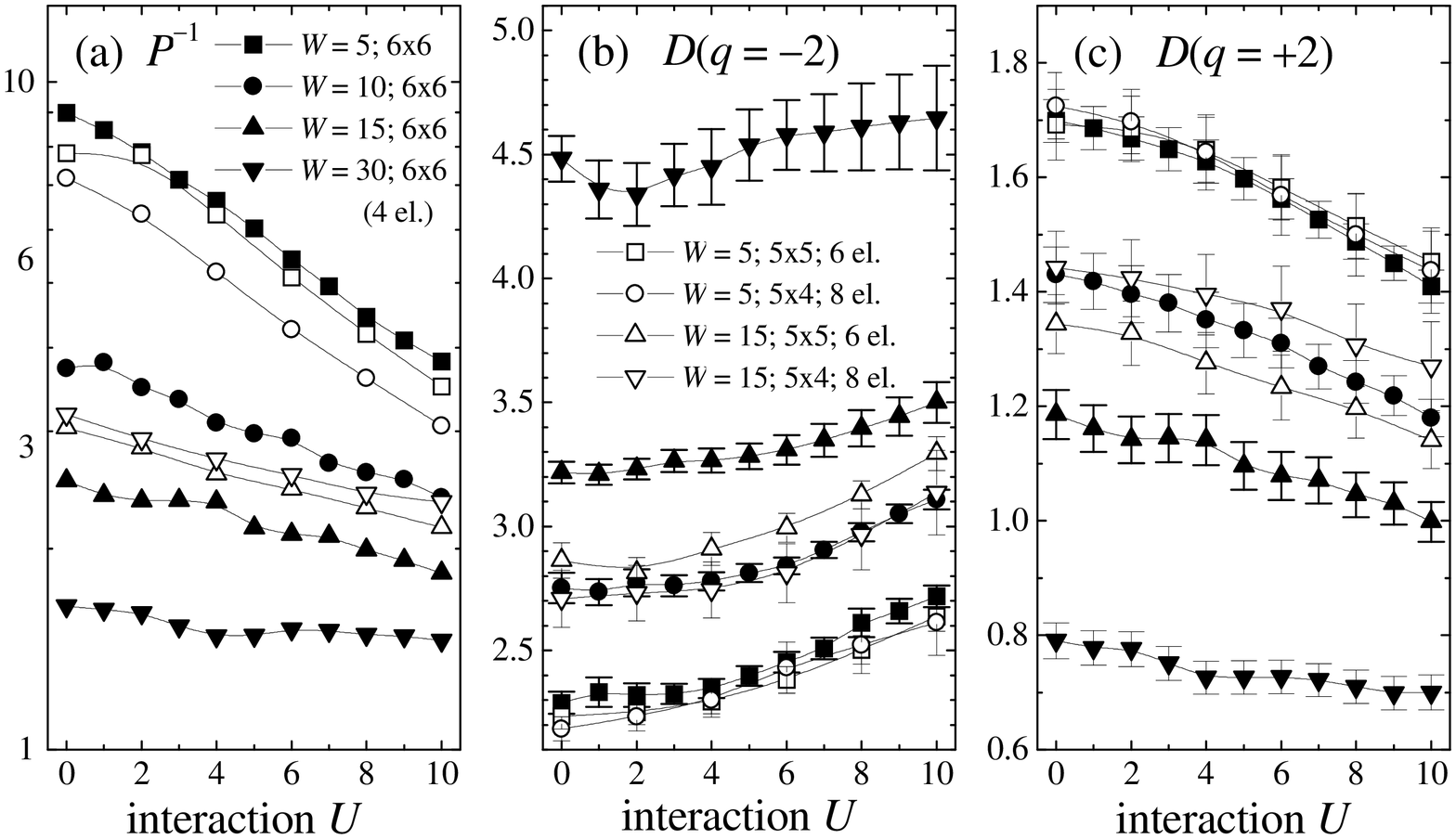}
\parbox{8.5cm}{\caption[]{\small 
(a) Participation ratio $P^{-1}$ as function of interaction
strength $U$ for $6 \times 6$, $4$ electrons; $5 \times 5$, $6$ electrons;
$5 \times 4$, $8$ electrons; and
different disorder values (all averaged over $100$ realizations of disorder). 
Parts (b) and (c)
show the results $D(-2)$ and $D(+2)$ from a multifractal analysis.
The results indicate a monotonous crossover to stronger localized 
states with increasing interaction or disorder.} 
\label{fig:results2}}\end{figure}

A similar feature is exposed 
in the inverse participation ratio
$P^{-1}$ with $ P = \sum_{\vec r} \vert\phi(\vec r)\vert^4$.
Figure~\ref{fig:results2}(a) shows the monotonous 
decrease of $P^{-1}$ with increasing $U$ and 
disorder $W$.  Again, there is no sign of an intermediate metallic phase.

In order to further elucidate the 
characteristics of the effective tunneling amplitude
$\phi(\vec r)$ we consider its
multifractal pattern (for a review, see \cite{janssen}).
The multifractal
exponents $\tau(q)$ are obtained from
 the scaling behavior of the partition
function
\begin{equation}
  Z_q(\ell) = \sum_i [p_i(\ell)]^q \sim \ell^{\tau(q)},
\end{equation}
where the sum runs over all boxes $i$ of linear size $\ell$ and $p_i(\ell)$ 
are the probability densities in each box.  Figures~\ref{fig:results2}(b) and 
(c) show $D(q) = \tau(q) /(q-1)$ for $q=-2$ and $+2$ as function of
interaction strength $U$.  Again, we do not find any indication of a 2DMIT. 

\begin{figure}\centering 
\epsfxsize8.0cm\epsfbox{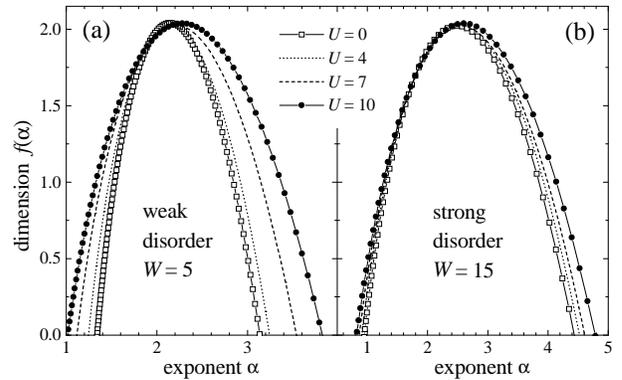}
\parbox{8.5cm}{\caption[]{\small 
Singularity spectra for effective multi particle transmission functions 
$\phi_{\vec r}$. For each singularity exponent $\alpha$, $f(\alpha)$ can be 
interpreted as the fractal dimension of the subset of boxes $i$, which 
are characterized by $p_i(\ell) \sim \ell^\alpha$. The spectra become wider
with increasing interaction or disorder indicating a crossover to
stronger localized states.} 
\label{fig:results3}}\end{figure}

Another way to present the results of the multifractal analysis is the 
singularity spectrum $f(\alpha)$, which is the Legendre transform of 
$\tau(q)$.  For each singularity exponent $\alpha$, $f(\alpha)$ can be 
interpreted as the fractal dimension of the subset of boxes $i$, which 
are characterized by $p_i(\ell) \sim \ell^\alpha$.  Figure~\ref{fig:results3}
shows the $f(\alpha)$ spectra for the $6 \times6$, $4$ electrons system at
two disorder strengths.  Since the 
singularity spectra are not independent of interaction strength (or
system size) the multifractality is not an indication of a critical
point.  The spectra become wider with increasing interaction or disorder
indicating stronger localized states, rather than 
generic multifractal states. Yet, the bottom line is that
we do not find any indication of a metallic phase but a mere
monotonous crossover into the regime of stronger localization.

In order to relate our results for the tunneling amplitude to the 
standard Kubo formulation of the conductance, we 
determine the 
distribution of the conductances $\sigma$ for different disorder 
configurations and, further, 
study the dependence of the average conductance on 
the interaction strength $U$.
The conductance at zero temperature is calculated using the 
many-particle Kubo formula \cite{berkovits96}
\begin{equation}
\sigma = \left({{\pi e^2}\over{M h}}\right)
\sum_{\alpha}^{'} {|\langle \alpha^N | J_x | 0^N \rangle|^2
\varepsilon_{\alpha,0}^N \gamma \over 
{{\varepsilon_{\alpha,0}^N}^2 + \gamma^2}},
\label{cond}
\end{equation}
where 
$\varepsilon_{\alpha,0}^N = \varepsilon_{\alpha}^N - \varepsilon_0^N$,
$J_x$ is the current operator and $M$ is the number of sites. 
The inelastic broadening $\gamma$ is chosen to be of the same order 
as the rescaled ``single electron'' level separation defined as
$B/N(M-N)$, where $B$ is the width of the many particle
energy band \cite{berkovits99,shepelyansky99}.
The calculation of the conductance is cumbersome since it 
requires the computation of many-body
eigenfunctions for all the low lying excitations (20 in
our calculation), which
is a difficult numerical task even 
within the Lanczos algorithm.

\begin{figure}\centering 
\epsfxsize8.5cm\epsfbox{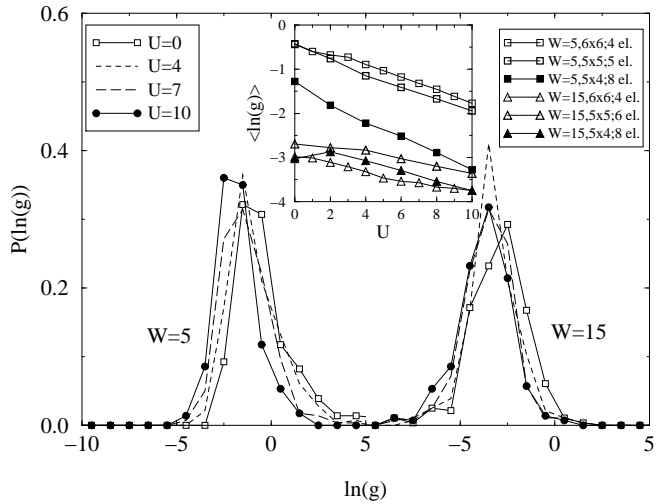}
\parbox{8.5cm}{\caption[]{\small 
The log of the dimensionless conductance distribution
for different values of disorder and interaction strength 
for the $6 \times 6$, $4$ electrons system. Inset: the 
average of the log of the dimensionless conductance as function of the 
interaction for the $6 \times 6$, $4$ electrons system 
(averaged over $280$ realizations of disorder); $5 \times 5$, $6$ electrons
($60$ realizations) and 
$5 \times 4$, $8$ electrons ($100$ realizations).}
\label{fig:cond}}\end{figure}

We first present the behavior of 
the conductance as function of $U$.
For the $6 \times 6$, $4$ electrons system,
the distributions of the dimensionless conductance $g = (h/e^2) \sigma$ 
are shown in Fig.~\ref{fig:cond} for two different values of disorder
($W=5,W=15$).  For the weaker disorder, which for this small system
corresponds in the non-interacting case to a marginal metal 
\cite{benenti99}, the distribution approaches
a log-normal distribution as interaction strength increases. 
For the stronger disorder, which even for the non-interacting case is in the 
localized regime, the log-normal distribution is preserved for any strength of
interaction. Since the conductance distribution of a localized system is
expected to follow a log-normal distribution \cite{shapiro86}, 
while a metallic system should follow a normal distribution
this behavior indicates that repulsive electron-electron
interactions suppress the metallic characteristics of 
the Kubo conductance. 
This is consistent with results reported sometime ago 
pertaining to the influence of {\em eei} on the 
conductance \cite{berkovits96,vojta98}.
In particular it has been shown that in the diffusive
regime the average conductance is suppressed by interactions, while only deep
in the localized regime, where the average current is very small,
some enhancement is possible\cite{vojta98}. 

The same trend can be seen when inspecting
the average logarithmic conductance, displayed in the inset of 
Fig.~\ref{fig:cond}.  Thus, neither the average conductance 
nor its distribution 
show any traces of a 2DMIT.

In conclusion, we have investigated the transport properties of spinless
electrons in disordered clusters. Both the tunneling amplitude and the 
conductance do not show any evidence of a 2DMIT. It has been shown that
the tunneling amplitude is the natural analog
of the single electron wave function 
appropriate for interacting many-particle systems.
This is quite attractive, since numerous methods developed to
connect the properties of single electron wave functions to transport 
properties of the system may now
be efficiently applied to the tunneling 
amplitude function.

RB and YA thank the Israel Science foundations centers of Excellence
Program and JK thanks the Minerva Foundation for financial support.

\end{multicols}

\vspace{-0.5cm}
\begin{thebibliography} {50}
\vspace{-1.5cm}

\bibitem{kravchenko94} S. V. Kravchenko et al., Phys. Rev. B {\bf 50}, 
8039 (1994); {\bf 51}, 7038 (1995);  
P. T. Coleridge et al., Phys. Rev. B {\bf 56}, R12764 (1997);
D. Popovic, A. B. Fowler, and S. Washburn, Phys. Rev. Lett. 79, 1543 (1997);
V. M. Pudalov et al., JETP Lett. {\bf 66}, 175 (1997);
Y. Hanein et al., Phys. Rev. Lett. {\bf 80}, 1288 (1998);
Y. Simmons et al., Phys. Rev. Lett. {\bf 80}, 1292 (1998);
J. Yoon, et al. Phys. Rev. Lett. {\bf 82}, 1744 (1999). 

\bibitem{hamilton99}
A. R. Hamilton, et al. Phys. Rev. Lett. {\bf 82}, 1542 (1999).

\bibitem{abrahams79}
E. Abrahams, et al., Phys. Rev. Lett. {\bf 42}, 673 (1979).

\bibitem{dobrosavljevic97}
V. Dobrosavljevic, et al., Phys. Rev. Lett. {\bf 79}, 455 (1997);
C. Castellani, C. DiCastro, and P. A. Lee, Phys. Rev. B {\bf 57}, 9381 (1998). 

\bibitem{talamantes96} J. Talamanes, M. Pollak, and L. Elam,
Europhys. Lett. {\bf 35}, 511 (1996).

\bibitem{cuvas99} E. Cuvas, Phys. Rev. Lett. {\bf 83}, 140 (1999).

\bibitem{benenti99} G. Benenti, X. Waintal, and J.-L. Pichard,
 Phys. Rev. Lett. {\bf 83}, 1826 (1999).

\bibitem{waintal99}
X. Waintal, G. Benenti, and J.-L. Pichard, Europhys. Lett. {\bf 49},
466 (2000).

\bibitem{shepelyansky99} D. L. Shepelyansky,  Phys. Rev. B {\bf 61}, 4588 
(2000); D. L. Shepelyansky and P. H. Song, Ann. Phys. (Leipzig) {\bf 8}, 
665 (1999).

\bibitem{denteneer99} P. J. H. Denteneer, R. T. Scalettar, and N. Trivedi,
Phys. Rev. Lett. {\bf 83}, 4610 (1999).

\bibitem{papadakis99}
S. J. Papadakis et al., Science {\bf 283}, 2056 (1999); 
Y. Yaish et al., (cond-mat/9904324);
M.Y. Simmons et. al, (cond-mat/9910368);
V. Senz et al., (cond-mat/9910228).

\bibitem{altshuler99} B. L. Altshuler and D. Maslov, Phys. Rev. Lett. 
{\bf 82}, 145 (1999); T. M. Klapwijk and S. Das Sarma, Solid State Commun.
{\bf 110}, 581 (1999); S. Das Sarma and  H. E. Hwang, Phys. Rev. Lett. 
{\bf 83}, 164 (1999).

\bibitem{edwards72} J. T. Edwards and D. J. Thouless, 
J. of Phys. C {\bf 5}, 807  27 (1972). 

\bibitem{berkovits96} 
R. Berkovits and Y. Avishai, Phys. Rev. Lett. {\bf 76}, 291 (1996).

\bibitem{akkermans97}
E. Akkermans and  J-.L.  Pichard, Euro. Phys. J. B , {\bf 1}, 223 (1997). 

\bibitem{shklovskii93}
B. I. Shklovskii et. al, Phys. Rev. B. {\bf47}, 11487 (1993).

\bibitem{berkovits94} R. Berkovits, Europhys. Lett. {\bf 25}, 681 (1994).

\bibitem{efros95} A. L. Efros and F. G. Pikus, Sol. Stat. Comm. {\bf 96},
183 (1995).

\bibitem{landauer57} R. Landauer, IBM J. Res. Dev. {\bf 1}, 233 (1957);
M. B\"uttiker, Phys. Rev. Lett. {\bf 57}, 1761 (1986).

\bibitem{meir92} Y. Meir and N. Wingreen, 
Phys. Rev. Lett.  {\bf 68}, 2512 (1992).

\bibitem{jeon99} G. S. Jeon et al., Phys. Rev. B {\bf 59}, 3033 (1999).

\bibitem{mahan90} G. D. Mahan, {\it Many-Particle Physics} 
(Plenum, New York, 1990).

\bibitem{berkovits98}
R. Berkovits and Y. Avishai, Phys. Rev. B. {\bf 57}, R15076 (1998).

\bibitem{janssen}
M. Janssen, Phys. Rep. {\bf 295}, 1 (1998).

\bibitem{berkovits99}
R. Berkovits, Phys. Rev. B. {\bf60}, 26 (1999).

\bibitem{shapiro86} B. Shapiro, Phys. Rev. B {\bf 34}, 4396 (1986).

\bibitem{vojta98} T. Vojta, F. Epperlein, and M. Schreiber, 
Phys. Rev. Lett. {\bf 81}, 4212 (1998).

\end{thebibliography}
\end{document}